\documentstyle[aps,preprint]{revtex}
\begin{document}
\draft
\title{Carbon Isotopes Near Drip Lines 
in the Relativistic Mean-Field Theory}
\author{M.M. Sharma, S. Mythili and A.R. Farhan}
\address{Physics Department, Kuwait University, Kuwait 13060}
\date{\today}
\maketitle
\begin{abstract}

We have investigated the ground-state properties of carbon isotopes 
in the framework of the relativistic mean-field (RMF) theory. 
RMF calculations have been performed with
the non-linear scalar self-coupling of the $\sigma$ meson using
an axially symmetric deformed configuration. We have also
introduced the vector self-coupling of the $\omega$ meson   
for the deformed mean-field calculations. The results show that
the RMF predictions on radii and deformations are in good agreement
with the available experimental data. It is shown that several
carbon isotopes possess a highly deformed shape akin to 
a superdeformation. The single-particle structure of nuclei away 
from the stability line has been discussed with a view to 
understand the properties near the neutron drip line.
Predictions of properties of carbon isotopes away from the
stability line are made. 

\end{abstract}
\vspace{1 cm}

\pacs{ }

\section{Introduction}

Radioactive beams are being used increasingly to produce nuclei
at the limits of the stability \cite{Rkl.92,MS.93,Tan.95,Ver.96,Proc.96}. 
Nuclei near the neutron and proton drip lines are becoming 
accessible to experiments. Abnormally large reaction cross-sections 
for unstable nuclei near drip lines have been interpreted as existence
of a large tail of neutron density in the exterior of nuclei.
Consequently, the so called halo of particles has been hypothesized 
for nuclei near drip lines. The case of $^{11}$Li has attracted a 
widespread attention \cite{Tan.orig}.
On one hand, properties of nuclei with halos and neutron skins are being
synthesized and studied experimentally, these properties are providing
a test bench to probe various theories and models, on the other hand. 
Description of very light nuclei in terms of a core and a set of
valence particles hovering around the core are proving to be generally
successful \cite{Zuk.}. For nuclei with a larger number of particles,
theories with an average field are being used increasingly. Thus, study
of nuclei with a large range of isospin puts various theories and
interactions to test their validity. At the same time, it should become
possible to discern the isospin dependence of the nuclear force by
studying nuclei at the extreme limits.

Relativistic Mean-Field (RMF) theory is one of the more successful
ones in recent times for describing nuclei with a large range of
isospin \cite{SW.86,Rei.89,Ser.92,Ring.96,SW.98}.
Earlier, the importance of the $\rho$-meson coupling and hence 
a proper asymmetry energy was emphasized by Sharma et al.
\cite{SNR.93}. Consequently, the RMF force NL-SH proved to be
very successful in describing various facets of nuclear structure.
Owing to a proper spin-orbit interaction in the RMF theory, it
became possible to describe the anomalous isotope shifts of Pb nuclei
\cite{SLR.93}. It was shown \cite{LS.95,LS.96,LFS.98}
that ground-state properties of nuclei
such as binding energies, charge and matter radii and deformation
properties of nuclei over a large part of the periodic table
are very well described within the RMF theory using the non-linear
scalar self-coupling of $\sigma$-meson. The Dirac-Lorentz structure 
of the nuclear force and the nuclear saturation based upon the attractive
component due to $\sigma$-meson and a repulsive component from
$\omega$-meson make the RMF theory an attractive tool to study
nuclear properties. An inherent spin-orbit interaction 
arising from the Dirac-Lorentz structure of nucleons gives rise to
appropriate shell effects. Such shell effects are duly responsible
to explain the behaviour of isotope shifts in Pb and other nuclear
chains \cite{SLR.93,LS.95}.  Based upon the isospin dependence of the 
spin-orbit interaction in the RMF theory, the Modified Skyrme Ansatz 
(MSkA) \cite{SLK.95} was proposed. Herein, the spin-orbit potential 
was proposed to contain only one-body contribution. The MSkA 
\cite{SLK.95} was shown to exhibit shell effects which are in accord with
the experimental data. Consequently, isotope shifts in Pb nuclei could 
be well reproduced with in the MSkA.

With the RMF theory having been developed to possess several
advantages over the non-relativistic theories, 
we study here the ground-state properties of the chain of carbon
isotopes. The carbon isotopes away from the stability line have
been the focus of experimental study whereby interaction cross-sections
for heavier carbon isotopes have been measured \cite{Lia.90}.
The matter and hence neutron radii of very neutron-rich isotopes have
thus been deduced. The nuclear structure of nuclei away from the
stability line and especially those in the vicinity of drip lines is
not yet fully understood. A few attempts have been made to explore
theoretically the ground-state properties of the carbon isotopes
\cite{Ren.96}. In the present work, we use the RMF theory to 
investigate systematically the nuclear structure of even-even
carbon isotopes from proton drip line to the neutron drip line.  
We employ the non-linear scalar potential of the $\sigma$-meson
as well as the model with the non-linear self-coupling of 
both the $\sigma$ and $\omega$ mesons. We will discuss the 
nuclear structure of carbon nuclei away from the stabiliy line.
A comparison will also be made with some results obtained
using the relativistic Hartree-Bogolieubov approach.
It will be interesting to see to what extent the mean-field
approach will be able to describe the properties of known light
nuclei such as carbon isotopes. In Section II, we present the 
formalism of the RMF theory and associated models of the $\sigma$ 
and $\omega$ meson potentials. 
The ensuing section provides details of the RMF calculations. 
In section IV we present the results and in the last section we 
will summarize our conclusions.

\section{Relativistic Mean-Field Theory}

The starting point of the RMF theory is a Lagrangian density \cite{SW.86}
where nucleons are described as Dirac spinors which interact via the
exchange of several mesons. The Lagrangian density can be written in
the following form:

\begin{eqnarray}
\nonumber
{\cal L} &=&\bar\psi (i\rlap{/}\partial -M) \psi +
\,{1\over2}\partial_\mu\sigma\partial^\mu\sigma-U(\sigma)
\nonumber\\
&&-{1\over4}\Omega_{\mu\nu}\Omega^{\mu\nu}
+{1\over2}m_\omega^2\omega_\mu\omega^\mu
+{1\over4}g_4(\omega_\mu\omega^\mu)^2
-{1\over4}{\bf R}_{\mu\nu}{\bf R}^{\mu\nu} 
+ {1\over2}m_{\rho}^{2}
 \mbox{\boldmath $\rho$}_{\mu}\mbox{\boldmath $\rho$}^{\mu}
-{1\over4}F_{\mu\nu}F^{\mu\nu} 
\nonumber\\
&&  -~g_{\sigma}\bar\psi \sigma \psi~
     -~g_{\omega}\bar\psi \rlap{/}\omega \psi~
     -~g_{\rho}  \bar\psi
      \rlap{/}\mbox{\boldmath $\rho$}
     \mbox{\boldmath $\tau$} \psi
     -~e \bar\psi \rlap{/}{\bf A} \psi. 
\label{eq1}
\end{eqnarray}
The meson fields included are the isoscalar $\sigma$ meson, the 
isoscalar-vector $\omega$ meson and the isovector-vector $\rho$ meson. 
The latter provides the necessary isospin asymmetry. The bold-faced 
letters indicate the isovector quantities. The model contains also a 
non-linear scalar self-interaction of the $\sigma$ meson :

\begin{equation}
U(\sigma)~={1\over2}m_{\sigma}^{2} \sigma^{2}~+~
{1\over3}g_{2}\sigma^{3}~+~{1\over4}g_{3}\sigma^{4} 
\label{eq2}
\end{equation}
The scalar potential (2) introduced by Boguta and Bodmer \cite{BB.77}
has been found to be useful for an appropriate description 
of surface properties,  although several variations of the 
non-linear $\sigma$ and $\omega$ fields have recently been 
proposed \cite{Furn.96}. We  have also included the vector
self-coupling  of the $\omega$-meson introduced by
Bodmer \cite{Bod.91}.
The corresponding term in the Langrangian is represented by the
coupling constant $g_4$. Here M, m$_{\sigma}$, m$_{\omega}$ and 
m$_{\rho}$ denote the nucleon-, the $\sigma$-, the $\omega$- and the 
$\rho$-meson masses, respectively, while g$_{\sigma}$, g$_{\omega}$, 
g$_{\rho}$ and e$^2$/4$\pi$ = 1/137 are the corresponding coupling 
constants for the mesons and the photon.

The field tensors of the vector mesons and of the electromagnetic
field take the following form:
\begin{equation}
\begin{array}{ll}
\Omega^{\mu\nu} =& \partial^{\mu}\omega^{\nu}-\partial^{\nu}\omega^{\mu}
\\
{\bf R}^{\mu\nu} =& \partial^{\mu}
                  \mbox{\boldmath $\rho$}^{\nu}
                  -\partial^{\nu} 
                  \mbox{\boldmath $\rho$}^{\mu}
                  -g_{\rho}(
                  \mbox{\boldmath $\rho$} \times
                  \mbox{\boldmath $\rho$})\label{eq3}\\
F^{\mu\nu} =& \partial^{\mu}{\bf A}^{\nu}-\partial^{\nu}{\bf A}^{\mu}
\end{array}
\end{equation}
The variational principle gives rise to the equations of motion. 
Our approach includes the time reversal and the charge conservation.
The Dirac equation can be written as:
\begin{equation}
\{ -i{\bf {\alpha}} \nabla + V({\bf r}) + \beta [ M + S({\bf r}) ] \}
~\psi_{i} = ~\epsilon_{i} \psi_{i}
\label{eq4}
\end{equation}
where $V({\bf r})$ represents the $vector$ potential:
\begin{equation}
V({\bf r}) = g_{\omega} \omega_{0}({\bf r}) + g_{\rho}\tau_{3} {\bf {\rho}}
_{0}({\bf r}) + e{1+\tau_{3} \over 2} {\bf A}_{0}({\bf r})
\label{eq5}
\end{equation}
and $S({\bf r})$ is the $scalar$ potential:
\begin{equation}
S({\bf r}) = g_{\sigma} \sigma({\bf r})
\label{eq6}
\end{equation}
the latter gives rise to the effective mass as:
\begin{equation}
M^{\ast}({\bf r}) = M + S({\bf r})
\label{eq7}
\end{equation}
The Klein-Gordon equations for the meson fields are time-independent
inhomogenous equations with the nucleon densities as sources.
\begin{equation}
\begin{array}{ll}
\nonumber
\{ -\Delta + m_{\sigma}^{2} \}\sigma({\bf r})
 =& -g_{\sigma}\rho_{s}({\bf r})
-g_{2}\sigma^{2}({\bf r})-g_{3}\sigma^{3}({\bf r})\\
\         \\
\  \{ -\Delta + m_{\omega}^{2} \} \omega_{0}({\bf r})
=& g_{\omega}\rho_{v}({\bf r}) -g_4 \omega_{0}^3({\bf r})  \\
\                            \\
\  \{ -\Delta + m_{\rho}^{2} \}\rho_{0}({\bf r})
=& g_{\rho} \rho_{3}({\bf r})\\
\                           \\
\  -\Delta A_{0}({\bf r}) = e\rho_{c}({\bf r})
\label{eq8}
\end{array}
\end{equation}
where $\omega_0 (r)$ and $\rho_0 (r)$ are the time-like components
of the $\omega$ and $\rho$ meson fields. The corresponding source 
terms are
\begin{equation}
\begin{array}{ll}
\nonumber
\rho_{s} =& \sum\limits_{i=1}^{A} \bar\psi_{i}~\psi_{i}\\
\             \\
\rho_{v} =& \sum\limits_{i=1}^{A} \psi^{+}_{i}~\psi_{i}\\
\             \\
\rho_{3} =& \sum\limits_{p=1}^{Z}\psi^{+}_{p}~\psi_{p}~-~
\sum\limits_{n=1}^{N} \psi^{+}_{n}~\psi_{n}\\
\                    \\
\ \rho_{c} =& \sum\limits_{p=1}^{Z} \psi^{+}_{p}~\psi_{p}
\label{eq9}
\end{array}
\end{equation}
where the sums are taken over the valence nucleons only. In the
the present approach we neglect the contributions of negative-energy 
states ($no-sea$ approximation), i.e. the vacuum is not polarized. 
The Dirac equation is solved using the oscillator expansion
method \cite{GRT.90}. 

The centre-of-mass correction to the total energy is included by taking
the centre-of-mass energy based upon the harmonic oscillator
prescription as given by
\begin{equation}
E_{cm} = {3\over4}\hbar\omega
\end{equation}

\section{\bf Details of Calculations}

The RMF calculations have been performed in a cylindrical
basis where axial symmetry has been maintained. The fermionic
and bosonic wavefunctions are expanded in a basis of harmonic
oscillator. For the expansion, we have taken 12 shells for the 
fermionic as well as bosonic wavefunctions. 

The pairing has been included using the BCS formalism. We have used 
constant pairing gaps which are obtained using the prescription 
of M\"oller and Nix \cite{MN.92} as given by

\begin{equation}
\begin{array}{ll}
\Delta_n = {4.8\over N^{1/3}} 
\\   
\\ 
\Delta_p = {4.8\over Z^{1/3}}
\label{e10}
\end{array}
\end{equation}

Only a few empirical data on the pairing gaps are available in this chain.
The pairing gaps obtained for a few nuclei from the experimental 
binding energies are found to be  consistent with the above prescription.
The centre-of-mass correction is included by using the zero-point
energy of a harmonic oscillator as in Ref. \cite{GRT.90}

We have used the RMF force NL-SH \cite{SNR.93} for the Lagrangian 
with the non-linear scalar coupling. This force has been employed widely 
to describe properties of several chains of nuclei. It is known to provide 
excellent results for nuclei on both the sides of the stability line. 
This force has been found to be especially useful for nuclei far away 
from the stability line.

For the Lagrangian with the vector self-coupling, we have used the 
force TM1 \cite{suga.94}. We have also used a newly
developed force NL-SV1 \cite{SHA.98} which
has been obtained by an exhaustive study of the ground-state
properties of nuclei within the framework of the non-linear
scalar and non-linear vector self-coupling. The details about
NL-SV1 will be provided elsewhere. The parameters and coupling
constants of the forces we have used are given in Table I.

\section{\bf Results and Discussion}

\subsection{Binding Energies} 
The binding energy of carbon isotopes in the deformed 
RMF calculations with the non-linear scalar force NL-SH and with the
non-linear scalar-vector forces TM1 and NL-SV1 are shown in 
Table II. The values correspond to the lowest energy for
the ground state. As carbon isotopes are perceived to be deformed, 
the RMF minimization was performed both for a prolate and an oblate
shape. The corresponding deformation for the lowest minimum and
a secondary minimum (if existent) will be discussed below. 
The binding energy for the secondary minimum are given
in parentheses. Experimental values from the mass
compilation of Audi and Wapstra \cite{AW.93} are also shown for 
comparison. 

It is seen that the force NL-SH describes the binding 
energies of light carbon isotopes well. For the heavier isotopes, 
the force NL-SH gives a slight overbinding. In comparison, The force 
TM1 gives an equally good description for the lighter nuclei. 
It is observed that NL-SH which is a force with a scalar
self-coupling of $\sigma$-meson only and TM1 which includes also the
self-coupling of $\omega$-meson produce binding energies which
are very close to each other for light as well as medium-heavy 
carbon nuclei. Whilst NL-SH lends an additional binding of 
about 1-2 MeV to the heavier carbon isotopes as compared
to the experimental values, TM1 provides a stronger overbinding to 
heavier carbon nuclei as compared to the NL-SH results as well as
the experimental values. 

The scalar-vector force NL-SV1, on the other hand, underestimates the 
binding of the very light carbon isotopes slightly. For the other
carbon isotopes, the agreement of the NL-SV1 binding energies with the
experimental values is qualitatively better than those of NL-SH and
TM1. A comparison of predictions of various forces with the experimental
data shows that the RMF theory provides a good description of
the binding energies of carbon isotopes. 

\subsection{The Quadrupole Deformation} 

The quadrupole deformation $\beta_2$ and the quadrupole moment
$Q_2$ obtained from the relativistic Hartree minimization with
various forces are shown in Fig 1. The  $\beta_2$ and Q$_2$
values are given in Table III. The deformation parameters
for NL-SH (Fig. 1.a) show that in the lowest-energy state the 
nucleus $^{10}$C is highly prolate ($\beta_2 \sim 0.54$). This 
nucleus also exhibits a secondary minimum with an oblate shape
($\beta_2 \sim -0.16$). The behaviour of $\beta_2$ using the forces
with the scalar-vector coupling is very different for this nucleus.
Both the forces TM1 (Fig. 1.b) and NL-SV1 (Fig. 1.c)
predict a well deformed oblate shape with $\beta_2 \sim -0.29$ and
$\beta_2 \sim -0.21$, respectively, for this nucleus. It can be seen
from Table III that a highly deformed (akin to superdeformation) secondary
minimum is also exhibited by this nucleus with TM1 ($\beta_2 \sim 0.64$)
and NL-SV1 ($\beta_2 \sim 0.58$).

The stable nucleus $^{12}$C is described as oblate shaped by all
the three forces. It is shown to be less deformed with NL-SH
as compared with TM1 and NL-SV1. TM1, in particular, gives this nucleus
a strogly deformed oblate shape ($\beta_2 \sim -0.39$). Whilst TM1
and NL-SV1 give a single well-defined oblate minimum for $^{12}$C,
NL-SH also predicts a spherical secondary minimum only about 100 keV
above the lowest minimum (see Table II). A comparison of the
$\beta_2$ values from RMF with the available experimental
information will be in order. The quadrupole deformation
for $^{12}$C has been estimated using various experimental probes.
Analysis of inelastic $\alpha$-scattering experiment \cite{Specht.71}
led to a value of $\beta_2 \sim -0.29$. However, another experiment
\cite{Yasue.83} with inelastic $\alpha$-scattering reported a
value of $\beta_2 = -0.40$. In comparison, a value of
$\beta_2 \sim -0.41$ has been deduced \cite{Simm.88} using inelastic
scattering of triton beam. Similar oblate deformations have also been
deduced from inelastic electron scattering \cite{Hori.71}.
All these experiments demonstrate unequivocally that $^{12}$C has
an oblate shape in its ground state. The RMF theory with both the scalar
self-coupling as well as with scalar-vector self-coupling models
describes the deformation of $^{12}$C very well. The deformations
predicted by the scalar-vector coupling models support
several experimental deductions.
 
The nucleus $^{14}$C with the magic neutron number N=8, and its
neighbour $^{16}$C are both shown to be spherical with all the three
forces. However, for nuclei above N=10, the influence of the magic
number N=8 diminishes rapidly and consequently the nuclei $^{18}$C,
$^{20}$C and $^{22}$C take up a well-deformed
oblate shape in RMF calculations with all the three forces.
It is seen that there is an abrupt onset of deformation as 2 neutrons
are added to the spherical nucleus $^{16}$C. The resulting deformation
for $^{18}$C is predicted to be less than that for ${20}$C for
any given force. The nucleus $^{20}$C with N=14 is well in the
middle of the shell where the largest value of $\beta_2$ is
obtained (see Table III). A further addition of a pair
of neutrons also brings about a well-deformed oblate shape for $^{22}$C.
The magnitude of $\beta_2$ for $^{22}$C is lower than that for
$^{20}$C. As the magic number N=20 approaches, nuclei
assume a spherical shape.

A comparative look at the values of $\beta_2$ and $Q_2$ 
amongst the three forces shows that for nuclei A=18-22, the force
NL-SH produces the lowest deformation, whereas TM1 is shown to provide
the largest value. The force NL-SV1, on the other hand, gives a value
of $\beta_2$ which lies between that of NL-SH and TM1.  Whilst TM1 and
NL-SV1 predict shape coexistence only for $^{10}$C, NL-SH is 
shown to give a secondary minimum with another shape for several 
nuclei.

A comment on the deformation properties of carbon isotopes in various
mean-field theories will be appropriate to make here. The mean-field
theories, in particular, those of the density-dependent Skyrme type
have generally given a spherical shape to the nucleus $^{12}$C. 
This nucleus is widely perceived to be oblate shaped. The RMF theory 
predicts such a shape and especially the scalar-vector model gives 
a highly deformed oblate quadrupole deformation for $^{12}$C.

\subsection{Neutron-Proton Deformations}

The difference in the quadrupole deformation of the neutron and proton
densities are shown in Fig. 2. The figure shows that for $^{10}$C,
the deformation for neutrons and protons is the same with NL-SH,
although the $\beta_2$ value is very large ($\sim 0.54$) in the lowest
energy state. The model with the scalar-vector self-coupling does show
a difference in the deformation of neutrons and protons. As discussed
above, forces TM1 and NL-SV1 with these models predict an oblate shape
for $^{12}$C. Both these forces also show the difference 
$\beta_n - \beta_p$ as positive. This implies that the proton field is
more deformed than the neutron field. TM1 predicts this difference 
($\sim 0.05$) to be slightly higher than NL-SV1. 

The $\beta_n - \beta_p$ is close to zero for $^{12}$C with all the
three forces although TM1 and NL-SV1 predict a stronger oblate shape
than NL-SH. The nuclei  $^{14}$C and $^{16}$C are spherical and hence
there is a vanishing $\beta_n - \beta_p$.
For the nuclei $^{18}$C, $^{20}$C and
$^{22}$C, a marked difference in the deformations of neutron and
proton fluids is seen. All these nuclei are neutron rich and predicted
to be oblate shaped with significantly large $\beta_2$ values. A large
negative value of $\beta_n - \beta_p$ for these nuclei means that
neutron deformation is considerably larger than the corresponding
proton deformation. The neutron number for these nuclei is 12, 14 and
16, respectively. The neutron deformation $\beta_n$ is accentuated
by these numbers being in the middle of the neutron shell.
The onset of the deformation at A=18 corresponds closely to a
significant occupation of the $\Omega^\pi = [202] 3/2^+$ orbital.
At the same time, there is a substantial lowering of this level as
A=18 (N=12) approaches. Details of single-particle structure will be
discussed in Section IV.E.

\subsection{\bf Nuclear Radii and Densities}

The $rms$ radii for neutron, charge and nuclear matter for the carbon 
nuclei are shown in Fig. 3. The values of radii are presented  
in Tables IV and V. The charge radius is obtained by folding 
the $rms$ proton radius with the finite proton size. Experimental
matter radii deduced from total reaction cross sections and neutron 
radii deduced therefrom \cite{Lia.90} are also given in Table IV. 
The radii shown in the figures correspond to the shape for the 
lowest energy state. 

The charge radius of carbon isotopes shows a roughly constant value
for most of the nuclei with all the forces. Only for $^{10}$C, 
there is a significant increase in the charge radius as compared to
$^{12}$C. This is due to the onset of proton drip line in going to
the more neutron-deficient isotopes. For isotopes heavier than
$^{12}$C, the charge radius shows only a marginal increase as
pairs of neutrons are added. Such a behaviour is common to most
isotopic chains. 

The charge radii predicted by the various RMF forces are given in
Table V. Charge radii of $^{12}$C and $^{14}$C from elastic electron 
scattering data \cite{Vries.87} are also shown in the table.
The experimental value of the charge radius is estimated 
at $2.47 \pm 0.02$ fm for $^{12}$C and at $2.56 \pm .05$ fm for 
$^{14}$C. In comparison the force NL-SH predicts the charge 
radius for $^{12}$C as 2.66 fm and both TM1 and NL-SV1 give it 
at about 2.7 fm. Thus, the RMF theory overestimates the charge radius
of $^{12}$C slightly. An explicit centre-of-mass correction might
improve the predicted value.

The $rms$ neutron radius shows a steady increase from A=10 to A=14
for all the forces (see Fig. 3). However, for nuclei above
A=14 there is a sudden increase in value.
The increase in the neutron radius is shown to be
phenomenal for $^{16}$C as compared to $^{14}$C in all the models.
Such an increase in the $rms$ neutron radius by $\sim$0.5 fm is
obviated by a major shell gap at N=8. The increase in the neutron
radius for the isotopes heavier than $^{16}$C is modest with 
a successive addition of a pair of neutrons. This increase in
the neutron $rms$ radius increases the neutron skin of the heavier
carbon isotopes. It is seen that the neutron radius of
$^{24}$C is only marginally larger than its lighter neighbour $^{22}$C. 
The magnitude of the increase in the $rms$ neutron radius of $^{24}$C or
of any of its neighbours is not so much as to characterize this as a 
neutron halo. It may be noted that $^{24}$C is predicted to be near
the neutron drip line in all the forces.  Thus, it is concluded that
the neuron halo is not present in the carbon isotopes near the
neutron drip line.

The matter radii as shown in Fig. 3 reflect much the behaviour of the
neutron radii. It is noteworthy that the features presented by 
various $rms$ radii for the carbon isotopes are similar for both the 
models with scalar self-coupling as well as with the scalar-vector 
self-coupling. There is also a similarity between the results of TM1 and
NL-SV1.

A comparison of the predictions of RMF matter and neutron radii
with the experimental values (Table IV) shows that the RMF values
describe the data very well. Predictions of all the forces are within the
error bars. Even for the heavier carbon isotopes such as $^{16}$C and 
$^{18}$C the RMF values show a very good agreement with the experimental
data. Thus, the RMF theory is able to provide reliable predictions
of the properties away from the stability line.

In Fig. 3a, we also show neutron radii taken from the spherical 
Relativistic Hartree-Bogoliubov (RHB) calculations \cite{Posc.97} 
using the force NL3. Herein, the finite-range Gogny pairing was
included. The RHB calculations take into account the 
effect of states near the continuum appropriately. As the force
NL3 is known to give results very close to those of NL-SH, a 
comparison of the NL3 results with those of NL-SH is worthwhile.
The comparison of the RHB neutron radii with the RMF+BCS results shows 
that the RHB values are very similar to the RMF+BCS ones. 
Especially, for spherical nuclei such as $^{16}$C and $^{24}$C, 
the agreement between our results and those of the RHB is remarkable. 
This suggests that the effect of the continuum on the carbon nuclei
near neutron drip line may not be significant. It is, however, not clear 
why RHB underestimates the neutron radius of $^{12}$C. 

The L=0 component of the vector ($\rho_v$) and scalar ($\rho_s$) 
neutron density for the carbon isotopes is shown in Fig. 4 for the
forces NL-SH, TM1 and NL-SV1. Both the vector and scalar densities 
show a similar behaviour. The scalar densities are slightly smaller
than the corresponding vector (baryonic) densities. The difference
in the vector and scalar densities is representative of the 
relativistic effects in the nuclear structure.

The increase in the radial extension of the densities is seen clearly
as the neutron number increases. The increase in the densities of
heavier carbon isotopes above $^{14}$C is mostly in the exterior
of nuclei, thus contributing to an increasingly large neutron skin
as more neutrons are addded. However, 
for the highly deformed carbon nuclei a considerable part of  
densities lies in the exterior of the quadrupole (L=2) component.
Figure 5 shows the L=2 component of the neutron vector density
$\rho_v$ obtained with the RMF forces for the three oblate deformed 
nuclei $^{18}$C, $^{20}$C and $^{22}$C. The negative density 
implies that the neutron matter is missing at the poles as compared 
to a spherical shape. The corresponding matter is accumulated along the 
equatorial plane.

As shown above the nuclei $^{18}$C, $^{20}$C and $^{22}$C are
predicted to be oblate shaped with all the three forces. As we have seen,
there are slight differences in predictions of the magnitude of the 
quadrupole deformation between various forces. The L=2 component of the 
density is sensitive to these differences in the deformation.
The L=2 densities show a negative peak which changes its position 
slightly about 2.5 fm from one nucleus to another. 

The densities given in Fig. 5a for the force NL-SH show that the
nucleus $^{18}$C, which has $\beta_2 \sim -0.32$, has the smallest
L=2 component as compared to $^{20}$C which has $\beta_2 \sim -0.41$.
For $^{22}$C the L=2 peak is smaller than that for $^{20}$C as 
$\beta_2$ for $^{22}$C is $\sim -0.31$ which is smaller than that
for $^{20}$C. However, the peak for $^{22}$C is bigger than
that for $^{18}$C due to a larger quadrupole moment for the latter.

A comparison of the L=2 densities for the various forces shows that
except for a slight variation in the depths of the peaks due to 
varying $\beta_2$ values and quadrupole moments, the features of the 
L=2 densities are very similar. This is primarily due to the reason 
that all the three forces considered here predict an overwhelmingly 
oblate shape for these nuclei and that the variation in the deformation 
in various models is similar from one nucleus to another.

We show in Fig. 6 the L=0 component of the proton vector density for
$^8$C, $^{10}$C and $^{12}$C. The lighter nuclei $^{10}$C and $^8$C
are proton rich and it is interesting to see how the proton density
changes as one approaches the proton drip line. Calculations with
the force NL-SH (Fig. 6a) show that the proton density in the 
interior of the nucleus $^{10}$C decreases as compared to that in
$^{12}$C. The decrease in the central density in interior of $^8$C
over its heavier counterparts is seen to be substantial. This takes place
at the expanse of the proton density at the surface. However, the spatial
extension of the proton density is not so much as to characterize it as a
proton halo. This is due to the Coulomb barrier which inhibits
formation of a proton halo. 

The behaviour of the proton density for the forces TM1 and NL-SV1
with the scalar-vector self-coupling is not very different from that
of NL-SH. Due to slightly different predictions of deformation for
$^{12}$C and $^{10}$C by these forces as compared to NL-SH 
(see Table III), densities of these two isotopes with scalar-vector forces
is only slightly different than with NL-SH. The behaviour of the density 
of $^8$C is similar with all the three forces. It may be pointed out 
that the densities of $^8$C are presented here only for a qualitative 
comparison.

\subsection{Single-Particle Levels}

In order to visualize the single-particle levels contributing to
the evolution of the properties of carbon isotopes as a function of
neutron number, we show the single-particle levels for NL-SH and
NL-SV1 in Figs. 7 and 8, respectively. The upper panel shows the
levels just below the continuum. In the lower panel, we show only a few
levels which are not so deep lying. The levels are identified by
the quantum number $\Omega^\pi$ consistent with the Nilsson scheme
\cite{Rag.95}. The numbers in the parentheses indicate the quantum numbers 
[Nn$_z\Lambda$]. Occupancy in the few highest lying $\Omega$ orbitals
is shown in Fig. 9 for NL-SH and NL-SV1. Here we show  
a comparison only between NL-SH and NL-SV1 which represent the
scalar self-coupling and scalar-vector self-coupling models,
respectively. The single-particle properties of the force TM1 are 
expected to be similar to that of NL-SV1. 

The lower panel of Fig. 7 shows that for the nuclei $^{10}$C and 
$^{12}$C which are predicted to be deformed (see Table III),
the level p$_{3/2}$ splits into [101]3/2$^-$ and 
[110]1/2$^-$. Th nucleus $^{10}$C being predicted as
highly prolate with NL-SH, the orbital [101]3/2$^-$
lies higher than [110]1/2$^-$. As one moves to $^{12}$C,
the two orbitals cross and the orbital [101]3/2$^-$
is suppressed in energy as a consequence of the oblate shape of this
nucleus as predicted by NL-SH. These two orbitals are degenerate
for $^{14}$C and $^{16}$C both of which are spherical. The degeneracy 
in the two orbitals is lifted again for the heavier carbon isotopes
$^{18}$C, $^{20}$C and $^{22}$C which are all predicted to be oblate
shaped. The splitting between the two $\Omega$ partners amounts to
about 10 MeV for these nuclei. Interestingly, the [110]1/2$^-$ 
orbital is lifted up significantly as compared to its degenerate 
(unperturbed) position.

The behaviour of the orbital [101]1/2$^-$ corresponding to the p3/2 level,
is seen to be influenced considerably by the presence of deformation. 
For the deformed nuclei and particularly those with an oblate shape, 
the orbital [101]1/2$^-$ is lowered significantly in energy as 
compared to the position in spherical case. For the prolate shaped 
nucleus $^{10}$C the opposite is true. 

For the force NL-SV1, the behaviour of the three orbitals below $-$10
MeV is very similar to that for NL-SH, as shown in the lower panel of
Fig. 8. As both $^{10}$C and $^{12}$C are predicted  to be oblate,
the crossing of [101]3/2$^-$ and [110]1/2$^-$ is not observed. 
Moreover, the splitting between these two levels is proportionate
to the $\beta_2$ value $-0.21$ and $-0.33$ for $^{10}$C and $^{12}$C,
respectively, as can be noticed readily from the figure. 

For nuclei up to A=14 (N=8), orbitals corresponding to the levels
1s1/2, 1p3/2 and 1p1/2 are filled successively. For nuclei above $^{14}$C,
the levels in the next shell start filling. The upper panel of Figs. 7
and 8 show the single-particle levels in this shell. These levels lie
in the neighbourhood of the Fermi surface which is shown by the dotted
line. The levels include $\Omega$ orbitals for 1d5/2, 2s1/2 and 1d3/2.
As the predictions of NL-SH and NL-SV1 on the deformations for $A > 16$
are similar, the single-particle structure in both Figs. 7 and 8 is
similar. 

The three $\Omega$ components for the 1d5/2 level are shown
by the dotted curve. These are degenerate for the spherical shape at
A=16. It is seen that as the deformed shape evolves for A = 18, 20 and 22
on adding neutrons to A = 16, the orbital [202]5/2$^+$ corresponding 
to 1d5/2 with the largest $\Omega$ value is suppressed. 
Part of this lowering is also expected from the increase in the depth 
of the potential well for nuclei richer in neutrons. The
counterparts with lower $\Omega$ values e.g. [211]3/2$^+$ and 
[220]1/2$^+$ are, on the other hand, lifted up in energy for the deformed
isotopes $^{18}$C, $^{20}$C and $^{22}$C. The three $\Omega$
components converge for $^{24}$C which is close to being spherical. 
The [200]1/2$^+$ component corresponding to 2s1/2 level is close to
being degenerate with the 1d5/2 level for the nucleus $^{16}$C.
The orbital [200]1/2$^+$ follows a mild lowering in energy with 
an increase in A. 

The $\Omega$ orbitals [202]3/2$^+$ and [211]1/2$^+$ for 1d3/2 level
also show a significant splitting for the deformed nuclei, the larger
$\Omega$ orbital being lowered in energy. The consequence of the
splitting between various $\Omega$ orbitals shown in the upper panel
of Figs. 7 and 8 is that there is a gap of about 3-4 MeV in the energy 
levels, which persists for A=18-22. 

Figures 7 and 8 show how the Fermi energy is changing with 
neutron number. For $^{16}$C, the Fermi energy decreases rapidly
as compared to $^{14}$C. This is due to the onset of the next shell
for neutrons. Above $^{16}$C, the Fermi energy decreases slowly
and approaches the vanishingly small value for $^{24}$C. This indicates
the onset of the neutron drip line. Such a behaviour is similar for
NL-SH as well as NL-SV1. 

We show in Fig. 9 the occupation probabilities for some of the orbitals
close to the Fermi energy.  The obitals are arranged in order of
increasing energy. Owing to the similarities in the predictions of the
deformations, the behaviour of the occupation numbers is similar
for NL-SH and NL-SV1 for the levels which are significant
only for the very neutron-rich carbon isotopes. The level [202]5/2$^+$
is filled up significantly already at A=18. The neighbouring levels
[200]1/2$^+$ and [202]3/2$^+$ (see Figs. 7 and 8) also show a
considerable occupancy at A=18 and are being filled up increasingly
up to A=22. The bunch of three levels [202]5/2$^+$, [200]1/2$^+$ and 
[202]3/2$^+$ is responsible for maintaining a very large 
oblate deformation for carbon isotopes with A = 18, 20 and 22.
The levels [211]1/2$^+$ and [211]3/2$^+$ play only a lesser role for
these nuclei. The latter is filled up only for A=24 which is predicted
to be spherical. The population of the orbital [202]3/2$^+$ 
$^{24}$C is quenched as this levels comes up again near the continuum
in its unperturbed position for the spherical shape.

In Fig. 10 we show the shapes of a few highly deformed oblate and
prolate configurations of carbon isotopes. The vertical axis in each 
figure is the axis of symmetry. The nucleus $^{10}$C has an oblate 
shape in the ground state as discussed above. This nucleus is also 
shown to give a secondary minimum with a highly prolate shape. 
Both NL-SV1 and TM1 forces give a similar predictions for the 
deformation. It is instructive to compare the single-particle structure
of $^{10}$C for the oblate and the prolate shapes.  The single-particle
structure for the two shapes is shown in Fig. 11. As there are only 4 
protons in $^{10}$C, two protons go inevitably to the 1s1/2 level.
The other two protons go to the second shell. For the oblate shape,
the $\Omega^\pi  = 3/2^-$ orbital is lower in energy and the splitting
between the $\Omega^\pi  = 3/2^-$ and $\Omega^\pi  = 1/2^-$ is small.
For the prolate shape, the lower $\Omega^\pi  = 1/2^-$ orbital is suppressed
in energy and contains a sizeable fraction of the nucleon occupancy.
The splitting between the two $\Omega^\pi$ orbitals is obviously larger 
for the case of the prolate shape (due to a larger value of $\beta_2$)
than the oblate case. The gap betwen the 1s1/2 level and the next shell 
is about 17.5 MeV for the prolate shape. The large deformation $\beta_2 =
0.59$ achieved using the force NL-SV1 and a larger value $\beta_2 = 0.64$
obtained for TM1 correspond to the so-called superdeformation of nuclei
whereby a 2:1 ratio of the axes are achieved within a harmonic oscillator
scheme. This also leads to a creation of new shells in the deformation
space \cite{Rag.95}. The oscillator frequencies $\omega_\perp:\omega_z$ for
the prolate shape are in the ratio 1.76 : 1, which lies in the vicinity of 
the ideal value 2 : 1 for the oscillator potential. Thus, $^{12}$C seems
to conform to a superdeformed configuration in the secondary minimum.

The nucleus $^{20}$C which is highly oblate deformed ($\beta_2 = -0.44$)
with the force TM1 is also shown in Fig. 10. This value of the oblate 
deformation is close to the 2:3 ratio  of the axes, whereby the axis of 
symmetry is accordingly shorter by this ratio. 

In Fig. 11 we compare single-particle levels for $^{10}$C for the
lowest energy state with those of the second minimum in energy for NL-SV1.
As the second minimum state is very close in energy (only $\sim 0.5$ MeV
above the lowest energy), it is instructive to see the difference in 
the single-particle structure of a shape-coexisting highly prolate 
and an oblate shape. For the highly prolate shape ($\beta_2 \sim 0.59$), 
both the $\Omega$ components of 1p3/2 level, i.e. [110]1/2$^-$ and 
[101]3/2$^-$ contribute to the prolate deformation, whereby the level
[110]1/2$^-$ plays a significant role. For the oblate shape, the orbital
[101]3/2$^-$ along with [101]1/2$^-$ of 1p1/2  contribute to
the deformation. Here the role of the level [110]1/2$^-$ for the oblate
shape becomes minimal. It is interesting to see a readjustment of various
$\Omega$ orbitals for two very different shapes but with almost the 
same total energy.

\section{Summary and Conclusions}

We have investigated the ground-state properties of carbon isotopes in
the framework of the relativistic mean-field theory using the 
self-couplings of $\sigma$ and $\omega$ mesons. Calculations
performed in the axially deformed configuration show that many 
carbon isotopes except those with magic neutron numbers are
significantly deformed. The force NL-SH with the scalar self-coupling
is shown to give a highly deformed prolate shape for $^{10}$C in the
lowest energy state. However, the forces NL-SV1 and TM1 with the
scalar and vector self-couplings predict an oblate shape for this nucleus. 
An oblate shape for $^{10}$C is also shown to occur with NL-SH,
however, in the secondary minimum. For $^{12}$C, the quadrupole 
deformation obtained with the various RMF forces is in good agreement 
with the values obtained from various experiments.

For all other isotopes above A=16, predictions on the deformation
of nuclei are similar in the scalar self-coupling and scalar-vector 
self-coupling models. Both the models predict an oblate shape for
the heavier carbon nuclei. The isotopes $^{18}$C, $^{20}$C and
$^{22}$C are predicted to be well deformed. The relative magnitude 
of the quadrupole deformations and quadrupole moments is largest for
TM1 and it is seen that deformations produced by NL-SH are 
slightly smaller than those of NL-SV1. It is observed that 
NL-SH exhibits the phenomenon of shape coexistence for several carbon 
isotopes whereas such a feature is presented by TM1 and NL-SV1 only 
for $^{10}$C where this nucleus is predicted to possess a very large 
prolate deformation in the secondary minimum lying only 0.5 MeV 
above the lowest enery state with NL-SV1 and 0.9 MeV above the lowest 
energy state with TM1. 

The relative quadrupole deformations of the neutron and proton mean
fields show that for lighter carbon isotopes, the neutron and proton
fields have almost comparable deformations in all the models. However,
isotopes above A=16 show significant differences in the
deformations of the neutron and proton mean fields. The quadrupole
deformation for neutrons are found to be larger than the corresponding
proton deformation for the heavier carbon isotopes. The magnitude of
the difference in the neutron and proton deformations is higher with
the forces with the scalar-vector self-coupling than with the scalar
self-coupling alone. 

The $rms$ matter and neutron radii obtained with the scalar and
scalar-vector self-coupling model in the RMF theory agree well 
with the experimental values on the matter and neutron radii
deduced from the total reaction cross-sections.
The neutron radii for
the carbon isotopes show a gradual increase with an increase in the
neutron number. Such a behaviour is predicted by all the forces
we have employed. However, the increase in the neutron radius and
consequently the radial extension of neutrons in space for very
neutron-rich carbon nuclei is not so much as to characterize it
as a neutron halo. This conclusion is in accord with the earlier
results obtained on carbon nuclei using spherical relativistic
Hartree-Bogoliubov calculations \cite{Posc.97}. A similar statement
can be made for carbon nuclei in the vicinity of the proton drip line
that the proton halo for these nuclei is suppressed.

The single-particle structure and the occupancy of levels near
Fermi surface show that in the midst of a large deformation for the nuclei
$^{18}$C, $^{22}$C and $^{22}$C, there is a pronounced gap in the shell
structure. We have also looked into the levels which contribute
significantly to a large value of deformation in these nuclei. 
It is shown that the magnitude of deformation in some carbon isotopes 
is akin to a superdeformation. The isotope $^{20}$C in its lowest energy
state and $^{10}$C in the secondary minimum exemplify this behaviour.
Such large deformations are produced predominantly in the scalar-vector
self-coupling model. These predictions on deformation properties 
change only a little from the force NL-SV1 to TM1.

\newpage

\bigskip

\begin{figure}
\caption{ The quadrupole deformation $\beta_2$ (upper) and the
the corresponding quadrupole moment $Q_2$ (lower) of the lowest energy 
ground-state in the RMF theory with (a) NL-SH (b) TM1 and (c) NL-SV1.
The $\beta_2$ and $Q_2$ (mb) values of the shape-coexistent secondary 
minimum are shown by a circle enclosed by a square. The $\beta_2$ values
for $^{12}$C from Ref.{\protect\cite{Specht.71}} (diamond), 
Ref. {\protect\cite{Yasue.83}} (triangle up) and from Ref. 
{\protect\cite{Simm.88}} (open circle) are also shown for comparison.}
\label{fig1}
\end{figure}

\begin{figure}
\caption{ The difference $\beta_n - \beta_p$ in the quadrupole 
deformations of the neutron and proton mean-fields. Points with 
a secondary minimum are shown by a square.}
\label{fig2}
\end{figure}

\begin{figure}
\caption{ The $rms$ neutron, matter and charge radii of C isotopes
in the RMF theory obtained with the force (a) NL-SH with the 
scalar self-coupling. The neutron radius from RHB calculations of
Ref. {\protect\cite{Posc.97}} is also shown. (b) Radii with TM1 and
(c) Radii with NL-SV1 with the scalar and vector self-coupling.}
\label{fig3}
\end{figure}

\begin{figure}
\caption{ The L=0 component of the neutron vector (baryonic) density 
$\rho_v$ (upper) and scalar density $\rho_s$ (lower) with (a) NL-SH
(b) TM1 (c) NL-SV1.}
\label{fig4}
\end{figure}

\begin{figure}
\caption{ The L=2 component of the neutron vector density
$\rho_v$ for a few deformed nuclei near neutron drip line 
with (a) NL-SH (b) TM1 and (c) NL-SV1.}
\label{fig5}
\end{figure}

\begin{figure}
\caption{ The L=0 component of the proton vector density 
$\rho_v$ compared for proton-rich nuclei with (a) NL-SH (b) TM1
and (c) NL-SV1.}
\label{fig6}
\end{figure}

\begin{figure}
\caption{ The neutron single-particle levels with the force NL-SH.
The lower panel shows the deeper lying $\Omega$ orbitals, whereas 
the upper panel shows levels in the vicinity of the Fermi surface.}
\label{fig7}
\end{figure}

\begin{figure}
\caption{ The neutron single-particle levels with the force NL-SV1.
The details are the same as for Fig. 7.}
\label{fig8}
\end{figure}

\begin{figure}
\caption{ The occupation probabilities for the highest five levels 
in the vicinity of the Fermi surface, for (a) NL-SH and (b) NL-SV1.}
\label{fig9}
\end{figure}

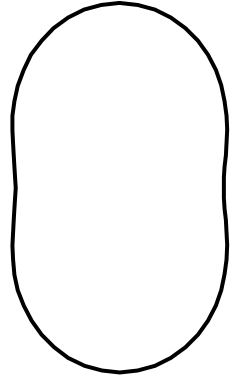
\begin{figure}
\caption{ The shapes of a few strongly deformed carbon isotopes. 
The upper panel shows the shape for the lowest energy state of
$^{10}$C and $^{20}$C. In the lower panel the prolate deformation
for the secondary minimum for $^{10}$C is shown. The vertical axis
represents the axis of symmetry.}
\label{fig10}
\end{figure}

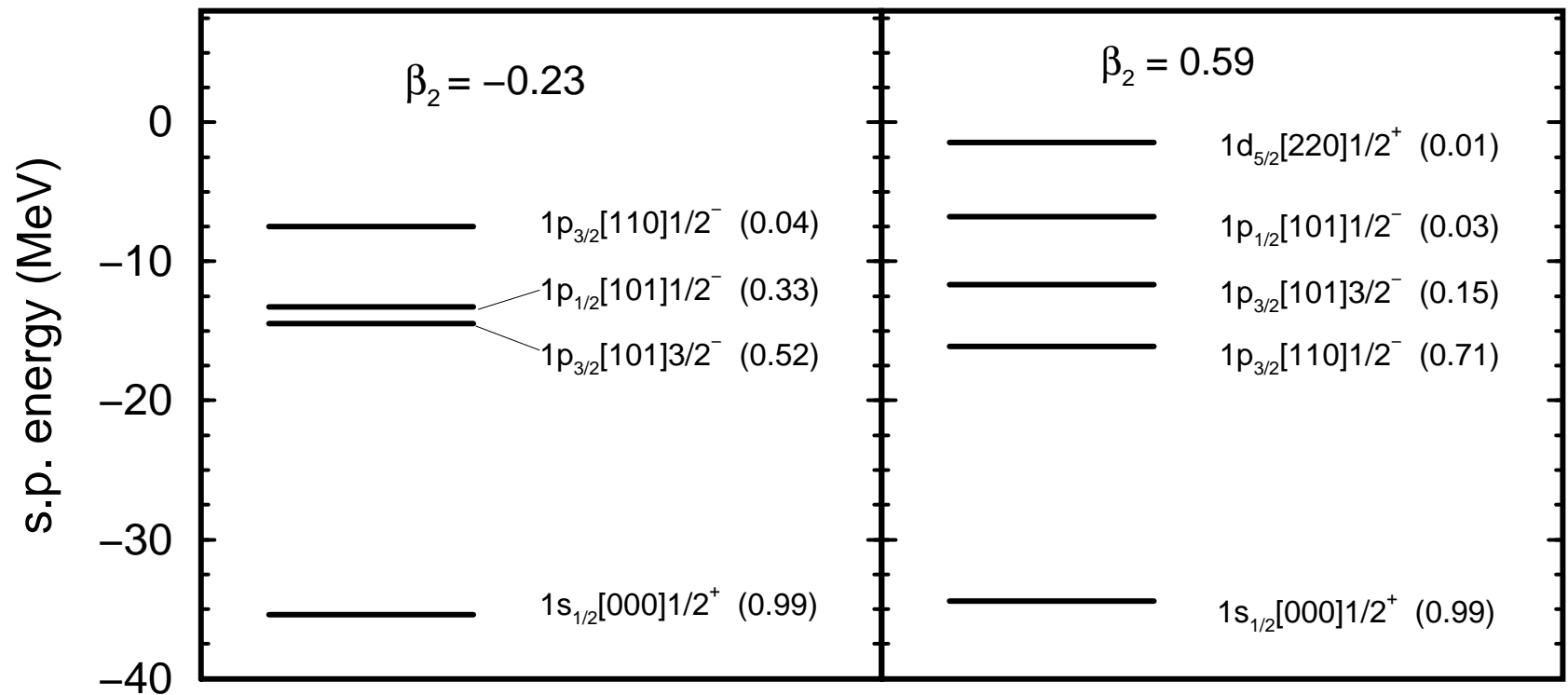
\begin{figure}
\caption{ The single-particle structure of the lowest energy state 
(oblate) and the highly deformed secondary minimum (prolate) 
for $^{10}$C using the force NL-SV1. The particle occupation numbers
are given in the parentheses.}
\label{fig11}
\end{figure}

\newpage
\widetext
\begin{table}
\caption{ The Lagrangian parameters of the forces NL-SH, TM1 and
NL-SV1 used in the RMF calculations.}
\begin{center}
\begin{tabular}{cccc}
   & NL-SH     &  TM1     &  NL-SV1     \\
\hline
M                 & 939.0     & 938.0    &  939.0      \\
$m_{\sigma}$      & 526.05921 & 511.198  &  510.03488  \\
$m_{\omega}$      & 783.0     & 783.0    &  783.       \\
$m_{\rho}$        & 763.0     & 770.0    &  763.       \\
$g_{\sigma}$      & 10.44355  &  10.0289 & 10.12479    \\
$g_{\omega}$      &12.9451    &  12.6139 & 12.72661    \\
$g_{\rho}$        &4.38281    &   4.6322 &  4.49197    \\
$g_{2}$           &-6.90992   &  -7.2325 & -9.24058    \\
$g_{3}$           &-15.83373  &   0.6183 &-15.388      \\
$g_{4}$           &0.0        &  71.5075 & 41.01023    \\
\end{tabular}
\end{center}
\label{Tab1}
\end{table}
\newpage
\noindent\begin{table} 
\vspace{-0.5cm}
\begin{center}
\caption{ The binding energies (in MeV) of even-even C isotopes 
obtained for the lowest energy state with the forces NL-SH, TM1
and NL-SV1. The empirical values (expt.) 
available are also shown for comparison. The numbers in the parantheses 
indicate the existence of a secondary minimum in the vicinity of the 
lowest-energy ground-state.}
\bigskip
\begin{tabular}{l l l l l l}
  &A &NL-SH           & TM1           & NL-SV1         &expt. \\
\hline\
 &10 &-60.4 (-59.8)   &-60.2 (-59.3)  &-57.9 (-57.4)   &-60.3 \\   
 &12 &-89.6 (-89.5)   &-90.1          &-87.6           &-92.2  \\
 &14 &-106.6          &-106.9         &-104.4          &-105.3 \\
 &16 &-112.3          &-112.4         &-109.6          &-110.8 \\
 &18 &-117.0 (-116.9) &-118.6         &-114.4          &-115.7 \\
 &20 &-121.8          &-123.4         &-118.5          &-119.2 \\  
 &22 &-122.7 (-122.4) &-123.9         &-118.6          &-120.3 \\ 
 &24 &-122.3          &-122.7         &-116.6          &       \\
\end{tabular}
\end{center}
\end{table} 

\newpage
\noindent\begin{table}
\vspace{-0.5cm}
\begin{center}
\caption{ The quadrupole deformation $\beta_{2}$ and the
quadrupole moment $Q_2$ (mb) obtained in the RMF theory for the
C isotopes.}
\bigskip
\begin{tabular}{l l l l l l l l l l}
  &   &       &     & $\beta_{2}$&  & &   $Q_{2}$ &    &\\
\hline
  & A & NL-SH & TM1 & NL-SV1 & & NL-SH& TM1& NL-SV1&\\
\hline\
  &10& 0.536 (-0.161)  & -0.294 (0.643) &-0.213 (0.584)& & 85.5 (-25.7)  
        & -46.9 (102.6) & -33.9 (93.1)  & \\
  &12& -0.23 (0.005)  & -0.388 &-0.328  & & -50.6 (0.9) & -83.9 & -70.9  & \\  
  &14& 0.000  & 0.011  & 0.005 & & 0.14  & 0.3   & 0.14    & \\
  &16& -0.005 &-0.006  & 0.004 & & -1.6  & -2.2  & 1.2    & \\ 
  &18& -0.316 (0.387)& -0.354 & -0.325 & & -134.3 (164.3)&-150.4 & -137.9 & \\ 
  &20& -0.405 &-0.444  & -0.418  & & -205.3&-224.7 & -211.6 & \\  
  &22& -0.308 (0.029) & -0.367 &-0.336 & & -183.2 (17.1)&-217.9 & -199.3 & \\ 
  &24& -0.006 & -0.16  &-0.07   & & -4.0  & -112.8& -49.4  & \\
\end{tabular}
\end{center}
\end{table}
\newpage
\noindent\begin{table}
\vspace{-0.5cm}
\begin{center}
\caption{ The $rms$ neutron radii $r_n$ and matter
radii $r_m$ (in fm) as obtained in the RMF theory for various RMF
forces. The experimental values {\protect\cite{Lia.90}} deduced 
from total reaction cross-sections are also shown for comparison}
\bigskip
\begin{tabular}{l c c c c c c c c c c c c l}
  &  & & $r_n$ & & & & &$r_m$& &\\
\hline
  & A & NL-SH& TM1 & NL-SV1 & expt. & &NL-SH&  TM1 & NL-SV1& expt. &\\
\hline\
  &10& 2.49 &2.49 & 2.48 &                 & &2.69 &2.69 &2.67 & & \\
  &12& 2.50 &2.57 & 2.56 & $2.49\pm0.16$ & &2.52 &2.59 &2.57 & 
  $2.48\pm0.08$& \\  
  &14& 2.57 &2.59 & 2.60 & $2.70\pm0.10$ & &2.54 &2.56 &2.56 &
  $2.62\pm0.06$& \\
  &16& 3.01 &3.03 & 3.04 & $2.89\pm0.09$ & &2.83 &2.85 &2.85 &
  $2.76\pm0.06$& \\ 
  &18& 3.18 &3.22 & 3.24 & $3.06\pm0.29$ & &2.98 &3.01 &3.02 &  
  $2.90\pm0.19$& \\
  &20& 3.32 &3.37 & 3.35 &                 & &3.11 &3.16 &3.13 & & \\  
  &22& 3.47 &3.52 & 3.54 &                 & &3.25 &3.30 &3.31 & & \\ 
  &24& 3.59 &3.63 & 3.66 &                 & &3.35 &3.40 &3.42 & & \\
\end{tabular}
\end{center}
\end{table}
\newpage
\noindent\begin{table}
\vspace{-0.5cm}
\begin{center}
\caption{ The charge radius $ r_c$ obtained with various RMF
forces. The experimental values for $^{12}$C and $^{14}$C from
the electron scattering data {\protect\cite{Vries.87}} are also shown.}
\bigskip
\begin{tabular}{c c c c c c c}
&  &   & $r_c$ &  & &\\
\hline
  & A & NL-SH& TM1 & NL-SV1 & expt. & \\
\hline\
  &10&2.93 &2.92 & 2.91  &               & \\ 
  &12&2.66 &2.73 & 2.71  & $2.47\pm0.02$ & \\
  &14&2.62 &2.64 & 2.63  & $2.56\pm0.05$ & \\
  &16&2.62 &2.64 & 2.63  &               & \\
  &18&2.64 &2.67 & 2.66  &               & \\
  &20&2.66 &2.7  & 2.68  &               & \\
  &22&2.67 &2.72 & 2.7   &               & \\
  &24&2.66 &2.72 & 2.69  &               & \\
\end{tabular}
\end{center}
\end{table}

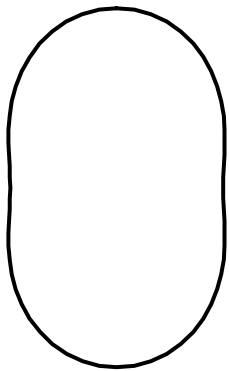
\end{document}